\documentclass{article}
\usepackage{caption}
\usepackage{amsmath}
\usepackage{amssymb}
\usepackage{lineno}

\newcommand{\bls}{\boldsymbol}
\newcommand{\bg}{\begin{linenomath*}\begin{eqnarray}}
\newcommand{\ed}{\end{eqnarray}\end{linenomath*}}
\newcommand{\eq}{=}

\begin{document}

\title{	Requiring negative probabilities from the thing researched, else that thing does not exist, is insufficient ground for any conclusion }

\author{
Han Geurdes,
Pensioner, 
\\\\
The Hague, Netherlands,
\\\\
2593NN, 164
\\\\
orcid 0000-0002-7487-1875
}





\maketitle 
\begin{abstract}
It is demonstrated that the statistical method of the famous Aspect - Bell experiment requires negative probability densities and negative 
probabilities from "the thing" researched, else that thing doesn't exist. The thing refers here to Einstein hidden variables.
This requirement
is absurd,
and therefore the results from such experiments are meaningless. 
\\
\\
\textbf{keywords:}{Statistical methodology $;$ Bell's theorem $;$ Einstein extra parameters$;$ Aspect's experiment }
\end{abstract}

\section{Introduction}
\subsection{Overview}
This overview is to help the reader understanding the simple nature of the statistical error in Aspect's experiment.
To begin with, 
Aspect's experiment is formulated in terms of classical probability. The possible model represented in Bell's formula, is estimated via the law of large numbers.

Subsequently, 
suppose that $\bls{a}$ is Alice's unit length parameter vector and $\bls{b}$ is Bob's. 
Let us define the angle between parameter vector $\bls{a}$ and $\bls{b}$. It is continuous variable $x=$angle$(\bls{a},\bls{b}) \in [0,2\pi)$.
Then it is esy to understand that
Aspect implicitly requires for a reproduction of the quantum correlation in experiment that $\cos(x)=P(x,\eq)-P(x,\neq)$.
In this way it is able to violate the CHSH inequality.
The, $P(x,=)=N(x,=)/N$ and $P(x,\neq)=N(x,\neq)/N$, while $N=N(x,=)+N(x,\neq)$. 
Furthermore the $N(x,=)=N(x,+,+)+N(x,-,-)$ with $+,+$ the indication of equal, $+$, polarization Alice, Bob measurement, etc.
And, we have $N(x,\neq)=N(x,+,-)+N(x,-,+)$
etcetera.

Then, with, $P(x,\eq)+P(x,\neq)=1$ and the simple trigonometry, $\cos(x)=1-2\sin^2(x/2)$, it can be derived that $P(x,\neq)=\sin^2(x/2)$ for $x\in[0,2\pi)$.
This, however, is impossible for any classical probability data.
It is because $F(x)=\sin^2(x/2)$ for $x\in[0,2\pi)$ isn't a probability function. 
The most easy way to see that is to look at the associated probability density $f(x)=dF(x)/dx$.
Therefore, the associated probability density is $f(x)=(1/2)\sin(x)$. This is not positive definite for $x\in[0,2\pi)$.

Aspect requires classical probability Einstein data to have a not positive definite probability density. 
No data exists with this property.
Einstein data is excluded beforehand. 

In order to meet possible objections with which people would possibly want to defend the design of the experiment: There is \emph{no} symmetry in the selection of angles. 
Such symmetry would require an overseer during the time of the experiment. 
Such overseer introduces non-locality into the experiment.
For instance, when Alice and Bob are at a distance that lies outside their light-cones, the overseer would be a non-local element in the design if the vectors $\bls{a}$ and $\bls{b}$ are known to the overseer.
Einstein's explanation is then excluded beforehand without the need of gathering data at all. 
Obviously,  such overseer makes the experiment worthless. 

Note too that after the fact introducing symmetry is equal to analyzing an experiment that wasn't performed. 
It is smuggling in a non-local element into the analysis and to pretend that it is a neutral transformation of the angle x.

If, for instance, an angle $x=7\pi/4$ 
is obtained, e.g. via clockwise rotation from random generated vector $\bls{a}$ to random generated vector $\bls{b}$.
Then an overseer is needed to make it a counterclockwise rotation so that $x=\pi/4$ can be used in the analysis.
We already discussed that during the experiment such is introducing a non-local element in the design of the experiment. 
After the end of the experiment employing such symmetry is therefore pretending that such an overseer was active during the experiment. 
The analysis is in that case directed at an experiment that wasn't performed and/or contains non-local elements. 

The conclusion is therefore: 
The range  of angle $x$  is $[0,2\pi)$ during the experiment. 
The angle must be analyzed accordingly after the end of the experiment. 
Otherwise, the experiment and its analysis both are meaningless. 

In addition, as was already stated previously.  
In the case of $x \in [0,2\pi)$, negative probabilities arise. 
Little wonder that Einstein positive data also isn't found in Aspect's experiment where 
$x \in [0,2\pi)$. 

Hence, Einstein data is openly or tacitly excluded despite all the careful considerations. 
The simple truth is that one can not violate statistical principles in an experiment and expect a serious conclusion about the data.
In the next sections more details will be given.

\subsection{Probability models}
The probabilistic basis of Bell's formula is the hypothesis that in nature, there "exists" a probability space, reference \cite{Ros}, 
$\left(
\Lambda,\Sigma,\mu
\right)
$
that co-determines the outcome of the experiment with polarization entangled photons. 
The $\Lambda$ here is the set of possible Einstein hidden extra parameters, 
$\lambda \in\Lambda$. 
The $\Sigma$ is the sigma-algebra set of  subsets of $\Lambda$. 
It represents the events. The $\mu$ represents the probability measure i.e., for all $S\in \Sigma$, we have, $0 \leq \mu(S) \leq 1$.
This probability space is one of many possible probability spaces. Next to the multitude of possible 
$
A(\bls{a},\lambda)
$
and 
$
B(\bls{b},\lambda)
$, 
this reflects the generality of Bell's formula. 
The particular probability space, together with the measurement functions, 
$
A(\bls{a},\lambda)
$
and 
$
B(\bls{b},\lambda)
$
is assumed to
reproduce the quantum correlation, $(\bls{a}\cdot\bls{b})=a_1b_1 + a_2b_2 + a_3b_3$.
The $\bls{a}$ and $\bls{b}$ are the unit length setting parameter vectors of the measurement instruments of the observers Alice and Bob.
Given that $x\in [0,2\pi)$, is the angle between  $\bls{a}$ and $\bls{b}$, then, $E(\bls{a},\bls{b})=\cos(x)$.
The angle is measured e.g. from $\bls{a}$ towards $\bls{b}$.
The Einstein locality conditions are such that Alice and Bob are free in their selection of their parameter vectors $\bls{a}$ and $\bls{b}$. 
And that, throuhgout the experiment, the setting of Alice's $\bls{a}$ is unknown to Bob and vice versa. 

In addition, the experiment must translate the possibility to find such a  probability space 
$
\left(
\Lambda,\Sigma,\mu
\right)
$
under Einstein locality conditions for 
$
A(\bls{a},\lambda)
$
and 
$
B(\bls{b},\lambda)
$ 
into an experiment related to probability space
$
\left(
U,\Phi,P
\right)
$.
If the locality conditions are observed in the experiment, also in the 
$
\left(
U,\Phi,P
\right)
$,
then a hypothetical probability space 
$
\left(
\Lambda,\Sigma,\mu
\right)
$
complies to the Einstein locality conditions as well. 

In Bell's correlation formula, there are no extra conditions present for 
$\lambda\in\Lambda$ to ensure \emph{local} hidden parameters. 
From an experimental point of view, locality is to be found \emph{only} in the not hidden parameters, i.e., the $\bls{a}$ and $\bls{b}$, together with the physical set-up, of the experiment.
Otherwise, Bell's correlation formula can not be applied. 
And, in that case, the inequalities derived from Bell's formula are also not applicable anymore.
Statistical analysis may, therefore, assume locality observed in the experiment. 
Furthermore,  in the experiment, the law of large numbers is employed to estimate a possible probability measure $P$, in the context of observable $x \in U$, when $U=[0,2\pi)$.
Hence, 
$
\left(
U,\Phi,P
\right)
$ is classical probability. 
The correlation is subsequently derived from the $P$.
Here we will consider $P$ further in the light of the Kolmogorovian axioms. 
\subsection*{1.3 Kolmogorov axioms}
Let us present some relevant Kolmogorovian axioms reference \cite{Kol} that are the foundation of  probability laws. 
A probability is a function of a set. We have three relevant axioms, 
\begin{itemize}
\item{
Positivity:
A probability, $P$, is \emph{never} negative
}
\item{
Certain event:
The probability $P$ of the universe set $U$ is unity 
}
\item{
Additivity:
If sets $Q$ and $R$ are disjoint, then $P(Q \cup R)=P(Q)+P(R)$.
}
\end{itemize}
Here, $Q$ and $R$ and $Q \cup R$ are subsets of $U$. Sets are connected to events via an appropriate random variable. 

\subsection*{1.4 Aspect's experiment }
With Bell's formula for correlation in reference \cite{B}, Einstein hidden parameters \cite{E} are modelled as classical probability random variables. 
The value(s) of those random variables, in a universe set $\Lambda$, are $\lambda$. 
The density is $\rho(\lambda)\geq 0$ for all $\lambda \in\Lambda$. 
The density is normalized, i.e., $\int_{\lambda\in\Lambda} \rho(\lambda) d\lambda = 1$.
Bell's correlation formula 
\bg\label{ed1}
E(\bls{a},\bls{b})=\int_{\lambda\in\Lambda} \rho(\lambda) A(\bls{a},\lambda)B(\bls{b},\lambda) d\lambda
\ed
is therefore surely embedded in Kolmogorovian classical probability. 
If one argues that a density somehow is associated with quantum variables, then $E$ in equation (\ref{ed1}), based on complex $\lambda$, may theoretically also violate the quantum result.
See however also reference \cite{Ge}, where a classical probability based Bell correlation is demonstrated to give the quantum correlation. 
In addition in the early days of quantum mechanics, Einstein already noticed the difference between classical probability, leading to Wien's law, and a kind of quantum probability leading to Planck's law  references \cite{How} and \cite{Kuhn}.
Moreover, there is the problem of an association of a quantum equation (\ref{ed1}) with an experiment, ruled by definition by classical probability laws. 
Finally, in the experiment use is made of classical probability based raw product moment correlation and the law of large numbers in order to obtain results without explicit knowledge of the probability density that resides behind the hidden variables $\lambda$.
Discrete densities, $\rho(\lambda)$, can be obtained with the use of Dirac delta functions.

In Bell's formula equation (\ref{ed1}), $\bls{a}$ is the unit length setting parameter vector of Alice's instrument and $\bls{b}$ is the unit length setting parameter vector of Bob's instrument.
Einstein's locality on the level of setting parameter vectors, is translated into: 
Alice doesn't know Bob's setting, and Bob doesn't know Alice's setting.

In the correlation we have,
$A(\bls{a},\lambda) \in \{-1,1\}$ Alice's spin or polarization measurement function $B(\bls{b},\lambda) \in \{-1,1\}$ Bob's spin or polarization measurement function.
Einstein's independence requirement is also obtained in a sufficiently large distance (see reference \cite{E}) between Alice and Bob.

For photons, it is sufficient to consider the angle $x=\angle(\bls{a},\bls{b})$ between the vectors $\bls{a}$ and $\bls{b}$.
This angle is a continuous variable in $0\leq x < 2\pi$ and is determined in the plane orthogonal to the direction of propagation.
The expectation value $E(\bls{a},\bls{b})$ in equation (\ref{ed1}) then reduces to $E(x)$.

In experiment, use is made of what can be called a raw product moment (rpm) correlation.
One can find it, e.g., in Aspects paper reference \cite{AGR}.
This rpm correlation is, of course, embedded in classical probability theory. 

Furthermore,  an excellent example of an experiment can be found in the literature, e.g., reference \cite{GWeiss}.
The rpm correlation for photons, called the raw product moment correlation, is,
\bg\label{ed2}
R(x)=\frac{N(x,\eq)-N(x,\neq)}{N(x,\eq)+N(x,\neq)}
\ed 
With, $N=N(x,\eq)+N(x,\neq)$ the total number of entangled photon pairs measured under angle x.
Here, given $0\leq x < 2\pi$,
$N(x,\neq)$ represents the number of unequal spin measurements by Alice and Bob, i.e., $(+,-),~ (-,+)$ and $N(x,\eq)$ represents the number of equal spin measurements by Alice and Bob, i.e., $(-,-),~(+,+)$.
This enables the rewriting $R(x)$ in equation (\ref{ed2}) as
\bg\label{ed3}
R(x)=1-2P(x,\neq)
\ed
Here, $P(x,\neq)$ equals $N(x,\neq)/N$.
It represents the (estimate) classical probability to find "$\neq$" spin under angle $x \in [0,2\pi)$. 
This is a statistical frequency of an event divided by a total, therefore, a probability estimate. 
The $N$ can be large if needed.

The set structure behind the probability space of the experiment is $(U,\Phi,P)$ with $U=[0,2\pi)$ and $\Phi$ the to $U$ associated sigma-algebra. The $P$ is the probability measure. 

The event "$x,\neq$" is represented by a random variable $X$. 
A random variable connects the probability set structure with what can be found in measurement (events).
Here, it associates the $\Phi$ set for the angle $x$, from the universe, $U=[0,2\pi)$ to the real numbers, i.e. $X:\Phi \rightarrow \mathbb{R}$. It does that in such a way that it connects a continuous variable to a set, viz. reference \cite{Lim}.
A probability is $P:\Phi \rightarrow  [0,1]$.

Therefore, a random variable $X$ is associated with the event "$\neq$" spin under angle $x$ and $x \in [0,2\pi)$.
The values on the real axis representing the event in the experiment in reference to the random variable $X$, of course, the $x\in[0,2\pi)$.

The reader can compare this to an experiment where the event "length of boys at a certain age $x$" is researched. The value of the random variable $X$ in this experiment is, of course, $x$.
One can ask if boys at age $x$ are more equal in length than at other ages. The $x$ is key here.
The probabilities $P(x,=)$ and $P(x,\neq)$ tell you something about a possible hypothesis.
Of course this is merely an example.

Similarly, the statistics of measurement of polarization denoted with $+$ or $-$, with $=$ pairs $(+,+)$ or $(-,-)$, and $\neq$ with $(+,-)$ or $(-,+)$ entangled photon pairs at the angle $x\in [0,2\pi)$.
The hypothesis, in this case, is expressed in equation (\ref{ed5}) below.
Furthermore, we are allowed here to assume ideal (no loss) polarization measurements.

The random variable $X$ is a continuous random variable.
The random variable $X$ attains continuous values because the angle between two vectors is a continuous variable.
It is therefore a mistake to think that $P(x,\neq)$ is the probability associated  with a discrete random variable. 
The continuous angle $x$ is part of the random variable $X$ describing the event $\neq$ given angle $x$. 
The point of coarse graining and discreteness is dealt with in a separate subsection. 

Subsequently, note that for a continuous random variable, the probability in a point is zero, see reference \cite{Hays}.
The probability for a continuous random variable is computed reference see \cite{Hogg} like, e.g., 
\bg\label{ed4}
P(0\leq X < x) = \int_0^x f(y) dy
=F(x)
\ed
The $f(y)$ here is obviously the probability density see e.g. \cite{Hogg}.
Furthermore, in Riemannian integration, the inclusion of limits doesn't make a difference in outcome.
\section*{2 Hypothesis}In the experiment, we ask if it is possible in principle that $R(x)$ can be equal to the quantum correlation $\cos(x)$. Because we have $\cos(x)=1-2\sin^2(x/2)$, 
this leads to the simple testing of the hypotheses
\bg\label{ed5}
H_0: P(0\leq X < x) = \sin^2(x/2)
\\\nonumber
H_1: \text{The hypothesis~~} H_0 \text{~~is false }
\ed
Therefore, note that the probability $P(x,\neq)$ is, in fact, $P(0\leq X < x)$. 
In this way, we can via the random variable $X$  have $\sin^2(x/2)$ associated to the set structure.
However, the following things immediately catch the eye.
\begin{itemize}
\item{
The function $F(x)=\sin^2(x/2)$ isn't monotone non-descending for $x \in [0,2\pi)$,
}
\item{
$P(0\leq X < 2\pi)=0$ instead of $1$,
}
\item{
The probability density, $dF(x)/dx=f(x)=(1/2)\sin(x)$ in equation (\ref{ed4}) is not positive definite for $x \in [0,2\pi)$.
}
\end{itemize}

\subsection*{2.1 Continuity \& negative probabilities}
With the fact that $F(x)=\sin^2(x/2)$ is not monotone non-descending for $x \in [0,2\pi)$, negative probabilities are required in order to let $P(0\leq X < x)$ meet $\sin^2(x/2)$.
Suppose, e.g., $S_1=[0,\pi)$ and $S_2=[\pi,3\pi/2)$. 
Then, $S_1\cap S_2 = \emptyset$ and when $\sin^2(x/2)$ is the probability function, $P(S_1 \cup S_2)=P([0,3\pi/2))=\sin^2(3\pi/4)=1/2$.
Because of the additivity in Kolmogorovian axioms, we also must have $P(S_1 \cup S_2) =P(S_1) + P(S_2)=1/2$. 
Again, when $\sin^2(x/2)$ is the probability function it follows, $P(S_1)=1$. 
But this leads, via $P(S_1)+P(S_2)=1/2$, to $P(S_2)=-1/2$ which is outside $[0,1]$.
Hence, $\sin^2(x/2)$ can not be a probability function for a continuous random variable. 
This is because negative probabilities arise from the additivity axiom 3 of Kolmogorov, presented above. 
This axiom is a basic part of probability theory, see also reference \cite{Hogg}.
\subsection* {2.2 Discreteness}
Special attention  again is given to the coarse graining and discreteness point above.
Here, we might use $P(x, \neq)$ in discrete points $x$.
With coarse graining and/or discreteness, the sum of discrete probabilities $\sin^2(x/2)$ over $x$ in nontrivial partitioning $\mathcal{X}$, cardinality e.g., $>5$ that contains $x=\pi$, always is larger than 1. 
The $\mathcal{X}$ is the discretisation of $U=[0,2\pi)$. 
There is a discrete sigma-algebra $\Phi'$ associated.
We have,
\bg\label{ed10}
\sum_{x\in \mathcal{X}\cup
\{\pi\}} \sin^2(x/2)>1
\ed 
This means there are sets $A$ with $P(A)>1$ and $A\in \Phi'$. For $P(U')$, we also might find $>1$.
For comparison, the discrete Poisson probability distribution is an example of a probability distribution of a discrete random variable with a summation to unity.
And please note that the Einstein hidden variables concept, \cite[page 320, ..unvolst{\"a}ndig..]{E}, wasn't rejected for a finite number of $x$. 
The claim was that for every possible $x\in[0,2\pi)$, we have no go Einstein variables.

\section*{Conclusion}
A classical probability is a function that projects a set into the interval $[0,1]$.
Aspect / Bell's experiment requires a Kolmogorovian probability to be not Kolmogorovian in order to meet the quantum correlation.  
The probability space 
$(\Lambda,\Sigma,\mu)$ is the backbone of a Bell correlation formula that might produce $\cos(x)$. 
The $\Lambda $ is the set of possible values $\lambda$ in the correlation formula. 
The $\Sigma$ is the set of subsets of $\Lambda$, which represents the events. It has a special structure called sigma-algebra. 
The $\mu$ refers to the probability measure translating the events into a probability of occurrence. 
In Bell's formula equation (\ref{ed1}), it is reflected in the density $\rho(\lambda)$.
Obviously, the $(\Lambda,\Sigma,\mu)$ that might produce $\cos(x)$ in the Bell correlation formula is completely unknown. 
Its existence in nature is tested with particular inequalities based on the correlation formula. 
Therefore, the law of large numbers, which is a part of classical probability theory, is employed to estimate the correlation in experiment viz. equation (\ref{ed2}). 

Note, in addition, we may only know the quantum world through the use of classical probability. 
Furthermore, no one looked "under the hood," so to peak and concluded: "no hidden variables" here. 
The non-existence of hidden parameters, therefore, follows from statistical methodology. 
From crooked statistical methodology, a sensible conclusion is not possible. 
The random variable $X$ for the event "$\neq$ given angle $x$" can not be pushed aside in a faithful translation of $(\Lambda,\Sigma,\mu)$ to an experiment. 
The value of $x$ is a decicive part of the events in the measurement. 

The points raised in the paper demonstrate that we will always find $H_1$ in equation (\ref{ed5}), in an experiment where the rpm correlation of equation (\ref{ed2}) is employed.
This is not because Einstein variables are impossible or that inequalities demonstrated that Einstein variables do not exist.
It is because it is \emph{not possible} in the data to find $H_0$ is true.
Data based on negative probability doesn't exist. 
For some initial research, please consult reference \cite{jqis}.

We conclude that the statistical methodology  of Bell's experiment  does not allow  any sensible conclusion about go or, no - go, Einstein extra local hidden parameters. 
An experiment is performed to find out if a state of affairs is found in the data or not.
If the statistical methodology requires the impossible of that state of affairs, i.e., the data gathering excludes giving $H_0$ true, then little wonder this state of affairs is not found.

\section*{Statements and Declarations }
\subsection*{Ethics}
Ethical Approval and Consent to Participate:
Not Applicable. 
\subsection*{Consent}
Consent for publication:
Not Applicable. 
\subsection*{Data}
Availability of supporting data: 
Not Applicable. 
\subsection*{Conflict of Interest}
No competing interests.
\subsection*{Authors' contribution}
Authors' contributions:
One author only. 
\subsection*{Funding}
Funding was not received.


\begin{thebibliography}{ll}
\bibitem{Ros} J.S. Rosenthal,
"Chapter 1", in
{A first look at rigorous probability theory}, 
Chapter 1, (Singapore: World Scientific 2006)
\bibitem{Kol} A.N. Kolmogorov, 
{
Foundations of the Theory of Probability}, pg. 1-3,NY: (Chelsea Publ. Com. 1950)
\bibitem{B} 
J.S. Bell,\emph{Physics} \textbf{1}, 195-204 (1964)
\bibitem{E}
A. Einstein, \emph{Dialectica}, 
\textbf{2}, 320-324 (1948) 
\bibitem {Ge}
H. Geurdes, K. Nagata and T. Nakamura,
\emph{ 
Russ. J. Phys. Chem.B.}
\textbf{ 15}, S68-S80 (2021)
\bibitem{How}
D. Howard, 
\emph{"Nicht sein kann was nicht sein darf", On the prehistory of EPR, 1909-1935: Einstein's early worries about the quantum mechanics of composite systems, in Sixty-two years of uncertainty: Historical, Philosophical, Physics inquiries into the foundations of quantum physics}, {A. Miller} Ed.,(NY: Plenum Publ., 1990)
\bibitem{Kuhn}
T.Kuhn, 
"Chapter 3", in,
{
Black-Body Theory and the Quantum Discontinuity, 1894–1912}, 
(USA Chicago: Univ of Chicago Press 1987) 
\bibitem{AGR}
A.Aspect, P.Grangier and G. Roger,
\emph{
Phys.Rev.Lett.}, \textbf{49}, 91-94 (1982)
\bibitem{GWeiss}
G. Weiss, T. Jennewein, C. Simon, H. Weinfurther and A. Zeilinger, 
\emph{
Phys. Rev. Lett.},\textbf{ 81}, 5093-5097 (1998)
\bibitem{Lim} J.Davidson,
{ Stochastic limit theory}, pg. 117,(Oxford: Oxford University Press 1994)
\bibitem{Hays}
W. Hays, 
{Statistics for the Social Sciences}, Chapter 3, (UK Plymouth: Holt, Reinhart and Winston Inc 1980)
\bibitem{Hogg} R.V. Hogg, and A.T. Graig,
{ Introduction to Mathematical Statistics, Third edition}, (UK Englewood Cliffs: Prentice-Hall 1995) 
\bibitem{jqis}
H. Geurdes, \emph{J. Quant. Inf. Sci.},
\textbf{13},177-182 (2023) ~and~~ https://arxiv.org/abs/2312.10038. 
\end{thebibliography}
\end{document}